\documentclass[prb,twocolumn,showpacs]{revtex4}
\usepackage{amssymb}
\usepackage{graphicx}
\usepackage{amsfonts}
\usepackage{amsmath}

\begin{document}

\title{Adiabatic quantum pumping, magnification effects and quantum size effects of spin-torque in magnetic tunnel junctions}

\author{A. Sorgente$^1$, F. Romeo$^1$, and R. Citro$^{1,2}$}
\affiliation{$^{1}$Dipartimento di Fisica ''E. R. Caianiello'' and
C.N.I.S.M., Universit{\`a} degli Studi di Salerno, Via Ponte don Melillo,
I-84084 Fisciano (Sa), Italy\\
Institute CNR-SPIN, Via Ponte don Melillo, I-84084 Fisciano (Sa), Italy}

\begin{abstract}

We study the adiabatic quantum pumping and quantum size effects of spin-torque in a magnetic
tunnel junction within a scattering matrix approach.
Quantum size effects are
predicted in the presence of a dc bias as a function of the thickness of the normal metal layer
inserted between two magnetic layers and of the fixed magnetic layer. In the presence of ac voltages,
the results for the spin-torque show a peculiar magnification effect and advantages of spin-torque pumping in actual devices are also discussed.
\end{abstract}

\pacs{73.23.-b,72.25.Pn,75.60.Jk,72.15.Qm}

\keywords{spin-torque, spin pumping, scattering matrix}

\maketitle

\section{Introduction}

Magnetic tunnel junctions (MTJs) are layered structures of
alternating magnetic layers (generally ferromagnetic) and
nonmagnetic layers (insulating or normal tunnel barriers) which
have recently attracted a lot of attention because of
magnetoresistance (MR) and spin-valve effects.\cite{1,2,3} MTJs
based on epitaxial MgO barriers\cite{4,5} are used in the magnetic
random access memory (MRAM) devices that work by spin-transfer
torque.\cite{6,7} While initially MTJs have attracted a
substantial attention for their tunnel magnetoresistance (TMR)
properties\cite{3,8,9,10},  more recently, the focus has shifted
to spin-transfer torque and current-induced magnetization
switching.\cite{11,12,13,14,15,16,17,18,19,20,21,22,23,24}
From the theoretical point of view, spin-transfer properties have
been studied extensively in spin valve structures based on various
model\cite{25,26,27,28,29,30} while for tunneling structures such
studies are still relatively few.\cite{15,17,18,19,20} The spin-transfer
torque has been analyzed in MTJs by first-principles electronic
structure calculations\cite{stiles02_1} or by Boltzmann
equation\cite{stiles02_2} and recently it has been revisited in
the Stoner model by scattering theory\cite{xiao08} and in
spin-valves by first principles with the aid of scattering
wavefunction\cite{wang08}. Ab initio studies of the spin torque in
metallic GMR junctions are also reported in
Ref.[\onlinecite{GMR_junctions}].
Despite the large amounts of spin-torque related papers, among the aspects
which require further attention are quantum size effects due to thin normal insertions layers and alternative ways of generating
magnetic torques.

In this paper we will analyze the spin-transfer torque in a quasi-one-dimensional
magnetic tunnel junction in which two magnetic regions (F1/F2) are separated by normal nonmagnetic spacers (NM).
We will focus on a quantum pumping mechanism as a nonconventional way of generating spin-torque and the advantages of such
mean compared to the conventional generation by external dc bias will be discussed. In particular, a magnification effect of the spin-torque will be predicted in the presence of ac bias. In the case of a conventional generation of spin-torque by an external dc bias, quantum size effects in the spin-transfer torque on a free layer will be analyzed and how to get information on the polarization at the interface will be illustrated.

For our analysis we choose a fully quantum mechanical treatment of
transport based on the scattering approach in a ballistic regime.
The motivation for concentrating on the ballistic regime, i.e. on
structures in which the transverse wavevector is conserved during
transport, derives from the evidence of quantum oscillations
observed in various FNIF structures. The ballistic regime is also
characterized by a spin-diffusion length $l_s$ and a
mean-free-path $l_m$ larger than the whole microstructure.

The organization of the paper is the following: In
Sec.\ref{sec:model} we introduce the model Hamiltonian and present
the scattering matrix approach generalized for the calculation of
the spin-torque in a MTJ. We then derive the expression of the torque components
caused by an external dc voltage bias and the spin torque pumped
via two ac voltages. In
Sec.\ref{sec:results} we present the results of the spin-torque for the structure shown in Fig.\ref{fig:fig1}.
Compared to Ref.[\onlinecite{romeo_2010}] we focus here on quantum effects related to the finite width of the magnetic layer and
on magnification effects of spin-torque by quantum pumping.

\section{ The model and formalism}
\label{sec:model}

Our system is shown in Fig.\ref{fig:fig1}, it is a multilayer
structure connected to two external leads in which two magnetic
layers are separated by nonmagnetic normal (NM) insertions (i.e. a NM/F1/NM/F2/NM
microstructure). For simplicity, one of the layers F1 is taken to have a width less
than the De Broglie wavelength and thus acts like delta barrier
spin-dependent potential. The system Hamiltonian is the following:
\begin{equation}
\label{eq:ham} H=-\frac{\hbar^2}{2m}\partial^2_x+\gamma_1(x)
\vec{n_1}\cdot \vec{\sigma}+\gamma_2(x) \vec{n_2}\cdot
\vec{\sigma}+ V(x),
\end{equation}
where $\gamma_{i=1,2}(x)=\frac{g\mu_B}{2}B_i(x)$ is the exchange
splitting, $\vec{n_i}$ is the unit vector in the direction of the
exchange splitting and $\vec{\sigma}=(\sigma_x,\sigma_y,\sigma_z)$
are the Pauli matrices. In our specific model $B_1(x)=B_1 \ell
\delta(x)$, while $B_2(x)=B_2 \chi(d_2,d_2+d_3,x)$ where the function $\chi$ is defined as $\chi(x_i,x_f,x)=\theta(x-x_i)\theta(x_f-x)$ and $\theta$ is the Heaviside step function. The spin independent scattering potential $V(x)$, which can be controlled by means of the gates G1/G2 is given by:
\begin{equation}
V(x)=\chi(-d_1,0,x)V_1+\chi(0,d_2,x)V_2+V_0 \ell \delta(x).
\end{equation}
The parameter $\ell$ which multiplies the local $\delta(x)$ potentials in $B_1(x)$ and $V(x)$ accounts for the layer finite size effects and is approximately equal to the size of the layer assumed much smaller than the Fermi wavelength\cite{note}. \\
We are interested here in the calculation of the spin-torque experienced by the layer F1 at $x=0$.

The spin torque $\vec{\mathcal{T}}$ is defined as time derivative
of the electron spin, represented by
the operator $\vec{s}$. This yields the total
spin torque as $\vec{\mathcal{T}} = -(i/\hbar) [\vec{s},H] =
\vec{\mathcal{T}}_1 +\vec{\mathcal{T}}_2$, where the
$\vec{\mathcal{T}}_j = 2 (\gamma_j/\hbar) \vec{n}_j \times \vec{s}$,
$j = 1,2$, is the torque experienced by the
j-th FM layer. However, when the magnetization direction of a
given layer is fixed (fixed layer) only the magnetization
direction of the other, the so-called free-layer, can be affected
by the local torque induced by the gradient of a
spin polarized current. Thus in the following we focus on the spin-torque on the free layer.
 This torque can be measured by tunnel magnetoresistance experiment. \\
 In our set-up of Fig.\ref{fig:fig1},
the layer $F1$ represents the free-layer, being $F2$ the fixed
layer whose magnetization direction is $\hat{n}_2=(0,0,1)$. The
torque generated on $F1$, i.e. $\vec{\mathcal{T}}_1$, lies on the
plane perpendicular to the magnetization direction
$\hat{n}_1=(\sin(\theta),0,\cos(\theta))$ of the free-layer since
$\hat{n}_1 \cdot \vec{\mathcal{T}}_1=0$.
The projection of $\vec{\mathcal{T}}_1$ parallel and perpendicular
to the free-layer can be expressed in terms of the following set
of basis vectors:
\begin{eqnarray}
\hat{\nu}_{\bot}&=&\frac{\hat{n}_2\times
\hat{n}_1}{|\hat{n}_2\times \hat{n}_1|}=\hat{y}\\\nonumber
\hat{\nu}_{||}&=&\frac{\hat{n}_1\times(\hat{n}_2\times
\hat{n}_1)}{|\hat{n}_1\times(\hat{n}_2\times
\hat{n}_1)|}=-\hat{x}\cos(\theta)+\hat{z}\sin(\theta),
\end{eqnarray}
where $\hat{x},\hat{y},\hat{z}$ are unit vectors along the direction of the cartesian axis.
The torque acting on the free-layer can thus be decomposed as
$\vec{\mathcal{T}}_1=\mathcal{T}^{||}_1\hat{\nu}_{||}+\mathcal{T}^{\bot}_1
\hat{\nu}_{\bot}$, where
\begin{eqnarray}
\label{eq:torque-pp}
\mathcal{T}^{||}_1
&=&\vec{\mathcal{T}}_1\cdot\hat{\nu}_{||}=\mathcal{T}_{1,y}\\\nonumber
\mathcal{T}^{\bot}_1
&=&\vec{\mathcal{T}}_1\cdot\hat{\nu}_{\bot}=-\mathcal{T}_{1,x}\cos(\theta)+\mathcal{T}_{1,z}\sin(\theta).
\end{eqnarray}
To calculate the torque components acting on the free-layer at
$x=0$ we make use of the scattering matrix approach and of the
following definition:
\begin{equation}
\label{eq:torque} \langle\mathcal{T}_{1,\mu}(x=0)\rangle
=\frac{2\gamma}{\hbar}\Bigl[\hat{n}(x=0)\times
\langle\vec{s}(x=0) \rangle\Bigl]_\mu ,
\end{equation}
where $\mu=\{x, y, z \}$, $\hat{n}(x=0)\equiv \hat{n}_1$, while
$\langle\vec{s}(x=0) \rangle$ represents the quantum average of the spin density operator on the free layer.

This quantum average can be evaluated by using the expression of the electron field operator
within the scattering approach of Ref.[\onlinecite{buttiker92}]:
\begin{eqnarray}
\label{eq:psi} &&\Psi_{\alpha}(x,t)=\sum_{\sigma}\int dE
\rho_{\alpha}(E) \exp\Bigl[-i\frac{E}{\hbar}t\Bigl]|\sigma
\rangle\times\\\nonumber &&[e^{i k x}a^{\alpha}_{\sigma}(E)+e^{-i
k x}b^{\alpha}_{\sigma}(E)],
\end{eqnarray}
where $\rho_{\alpha}(E)=[\sqrt{2\pi \hbar v_{\alpha}(E)}]^{-1}$ is
the density of states of the external lead $\alpha=1,2$, while $v_{\alpha}(E)$ is the
velocity of the electrons with wave vector $k(E)=\sqrt{2mE}/\hbar$.
The scattering operators $a^{\alpha}, b^{\alpha}$ for the incoming and
outgoing states, respectively, are related by the scattering
matrix $S$ through the relation $b^{\alpha}=\sum_{\beta}S^{\alpha
\beta}a^{\beta}$ and the notation  $a^{\alpha},b^{\alpha}$ stands for the following spinorial
representation:
\begin{equation}
\label{eq:spinorial} a^{\alpha}=\Bigl(\begin{array}{c}
                   a_{+}^{\alpha} \\
                   a_{-}^{\alpha}
                 \end{array}\Bigl),\\
\end{equation}
while $a^{\alpha \dag}=(a_{+}^{\alpha \dag}, a_{-}^{\alpha
\dag})$, and similarly for $b$. Using the relation $\langle
a_\sigma^{\alpha\dag}(E)a_{\sigma'}^{\beta}(E')\rangle=\delta_{\alpha \beta}
\delta_{\sigma,\sigma'}\delta(E-E') f_{\beta}(E)$, $f_{\beta}(E)$
being the Fermi function of the lead $\beta$, the spin density can be computed as the quantum average
$\langle \Psi^{\dag}_{\alpha}(x,t)\frac{\hbar}{2}\vec{\sigma}\Psi_{\alpha}(x,t)\rangle$. Since we are interested in the spin density on the free layer only the contribution from the closest lead, the left one ($\alpha=1$), can be retained.
Using this expression the spin torque can be evaluated via Eqs.(\ref{eq:torque-pp})-(\ref{eq:torque}).\\
Two different ways of generation of spin torque can be considered: one relies on the conventional application of dc external voltage bias $V$, the other is the adiabatic quantum pumping mechanism. The corresponding analytic expressions will be derived below.\\
Let us finally comment on a physical difference between the in-plane and out-of-plane spin torque components.
As shown they originate from different spin vector components of the spin current implying a further qualitative
differences between them. A natural difference
is that the out-of-plane component  is present in equilibrium, i.e. at zero bias, when the external leads present a spin dependent energy spectrum. This difference can also be
understood on general symmetry grounds ( see  Ref.[\onlinecite{stiles_prb79_2009}]).

\subsection{Spin torque by dc voltages}

The $\mu$-th
component of the spin density  $\langle s_{\mu} \rangle$ in
the approximation of a constant density of states at the Fermi level, $\rho(E)\sim \rho(E_F)$, can
be written as \cite{sharma-brouwer03-spin-curr}:
\begin{eqnarray}
\label{eq:spin-density} \langle s_{\mu}
\rangle\simeq\frac{1}{4\pi v_F}\sum_{\beta}\int dE Tr\{S^{1 \beta
\dag}(E)\sigma_{\mu}S^{1 \beta}(E) \}f_{\beta}(E),
\end{eqnarray}
$v_F$ being the Fermi velocity.
When a dc external voltage $V$ is applied to the microstructure $f_\beta(E)\rightarrow f_\beta(E\pm eV/2)$, and
Eq.(\ref{eq:spin-density}) can now be used to compute the torque components acting on the free-layer. To linear order in $V$ one gets:
\begin{eqnarray}
\label{eq:torque_par/perp_free-dc}
\mathcal{T}^{||}_1 &=& -\frac{w \Gamma}{2\pi}Tr[\sigma_yS^{12}S^{12\dag}]\\\nonumber
\mathcal{T}^{\bot}_1 &=& \frac{w \Gamma}{2\pi}Tr[(\sin(\theta)\sigma_z-\cos(\theta)\sigma_x)S^{12}S^{12\dag}],
\end{eqnarray}
where $w=eV/2$, while the dimensionless parameter $\Gamma$ is  $\Gamma=\frac{2m \gamma}{\hbar^2k_F}$.
Recalling that $\gamma=g\mu_B B_1 \ell/2$, $\Gamma$ can be rewritten as the Zeeman energy (normalized to the Fermi energy $E_F$) of the free-layer rescaled by the normalized effective length $k_F \ell$ (i.e. $\Gamma=(k_F\ell)\frac{g\mu_B B_1}{2E_F}$).
Apart the spin torque, one can define the torkance as the linear response to a small variation of the external bias $\delta w=e\delta V/2$,
$(\partial \mathcal{T}_\mu/\partial w) \delta w$, and its expression is then $T_\mu=\mathcal{T}_\mu/(eV/2)$.

\subsection{Quantum pumping of spin torque by ac external gates}
\label{sec:pumping}

\begin{figure}
\centering
\includegraphics[scale=0.4]{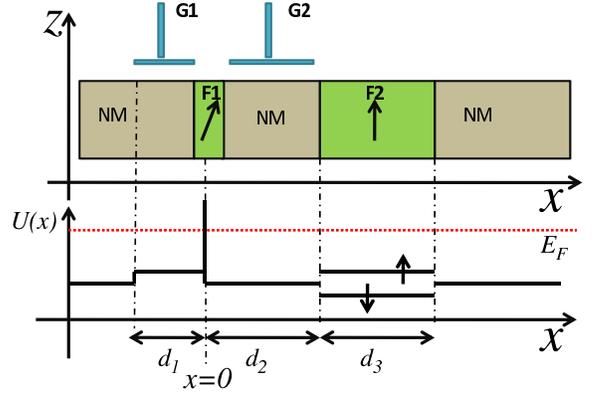}\\

\caption{Representation of the
NM/F1/NM/F2/NM system and of the respective potential energy. The spin current
flows along the $x$-direction, while the magnetizations $\hat{n}_1$ and $\hat{n}_2$
belong to the $x-z$ plane.} \label{fig:fig1}
\end{figure}

Quantum pumping\cite{thouless} is a well known quantum effect
for charges. In a charge quantum pump a dc particle current is
generated by the ac adiabatic modulation of at least two
out-of-phase independent parameters of the system in {\it absence of bias}. In
our calculation we will use the idea that in a magnetic layered
structure a pumping mechanism can generate spin currents other
than charge currents, and thus a spin-torque is generated on a
magnetic layer by the gradient of spin current, or equivalently by
the local spin-density (see Eq.(\ref{eq:torque})). Recently quantum pumping has been
proposed as an additional control of the spin flux in absence of
external dc bias\cite{wang-spin-curr-pump}.
Focusing on the microstructure of Fig.\ref{fig:fig1} and applying the idea of pumping, we modulate harmonically
in time the barriers heights by two top gates G1 and G2.
When the gates are varied adiabatically in time the
scattering matrix depends on time via the two varying external
parameters as $S(t)=S(X_1(t),X_2(t))$, where
$X_i(t)=X_i^0+X_i^{\omega}\sin(\omega t+\varphi_i)$ ($i=1,2$). In
the weak pumping regime, i.e. when $X_{1,2}^{\omega}\ll
X_{1,2}^{0}$, the scattering matrix can be expanded as follows:
\begin{equation}
\label{eq:exp_S} S^{\alpha \beta}(t)\simeq S_0^{\alpha
\beta}+\sum_{\eta= \pm1}S_{\eta}^{\alpha\beta}e^{i\eta \omega t},
\end{equation}
where $\omega=2\pi\nu$ is the pumping frequency and the matrices $S_{\eta}^{\alpha\beta}$ are given by
\begin{equation}
\label{eq:small_s}
S_{\eta}=-\frac{i\eta}{2}\Bigl[X_1^{\omega}(\partial_{X_1}S)_0+X_2^{\omega}e^{i\eta\varphi}(\partial_{X_2}S)_0\Bigl],
\end{equation}
$\varphi$ being $\varphi_2-\varphi_1$.
The Fourier transform of (\ref{eq:exp_S}) is then:
\begin{equation}
\label{eq:ft_exp_S} S^{\alpha \beta}(E)= 2\pi[ S_0^{\alpha
\beta}\delta(E)+\sum_{\eta=
\pm1}S_{\eta}^{\alpha\beta}\delta(E+\eta \omega)].
\end{equation}
Using this equation in evaluating the spin density,
its components per unit of area are given
by:
\begin{equation}
\langle s_\mu\rangle \simeq \frac{1}{4\pi v_F}\sum_{\eta \beta}\int
dE Tr\{S^{1\beta \dag}_{\eta}\sigma_{\mu}S^{1\beta
}_{\eta}\}f_{\beta}(E+\eta\omega).
\end{equation}
 Since no bias is present between the leads, i.e. $f_{1}(E)=f_2(E)=f(E)$,
and in the zero temperature limit, the $\mu$-th component of
spin-density per unit area to leading order in the adiabatic
frequency $\omega$ is:
\begin{equation}
\label{eq:spin-density-pump}
\langle s_\mu\rangle = -\frac{\hbar\omega}{4\pi v_F}\sum_{\beta\eta}\eta
Tr\{\sigma_{\mu}S^{1\beta}_{\eta}S^{1\beta
\dag}_{\eta}\}.
\end{equation}
Substituting (\ref{eq:spin-density-pump}) in
(\ref{eq:torque}) we obtain the explicit expressions of the pumped torque components acting on the
free-layer:
\begin{eqnarray}
\label{eq:torque_par/perp_free-ac} \mathcal{T}^{||}_1 &=&
\frac{\hbar \omega
\Gamma}{8\pi}X_1^{\omega}X_2^{\omega}\sin(\varphi)\sum_{\beta}Tr[A_y^{1\beta}+A_y^{1\beta\dag}]\\\nonumber
\mathcal{T}^{\bot}_1 &=& -\frac{\hbar \omega
\Gamma}{8\pi}X_1^{\omega}X_2^{\omega}\sin(\varphi)\sum_{\beta}Tr[\sin(\theta)(A_z^{1\beta}+A_z^{1\beta\dag})\\\nonumber
&-&\cos(\theta)(A_x^{1\beta}+A_x^{1\beta\dag})],
\end{eqnarray}
where the quantity
$A_{\mu}^{\alpha\beta}=i(\partial_{X_2}S^{\alpha\beta})_0^{\dag}\sigma_{\mu}(\partial_{X_1}S^{\alpha\beta})_0$ has been introduced.\\
The equivalent of the torkance in the dc case is obtained for the pumping case by $T_\mu=\mathcal{T}_\mu/(\hbar \nu/2)$.
In the following we introduce the dimensionless potential barriers
$r_i=\frac{V_i}{E_F}$ ($i=1, 2$) and $r_0=(k_F \ell)V_0/E_F$,
the normalized Zeeman energy of the fixed layer $h=\frac{g\mu_B B_2}{2 E_F}$ and the dimensionless distances $ k_F d_i$.
\section{Results}
\label{sec:results}
In the following we present the results for the torkance components per unit of area, or equivalently of the spin torque normalized by $eV/2$ in the dc case and by $\hbar \nu/2$
in the pumping case.
\subsection{dc case}
\begin{figure}[h]
\vspace{1.0cm}
\centering
\includegraphics[scale=0.8]{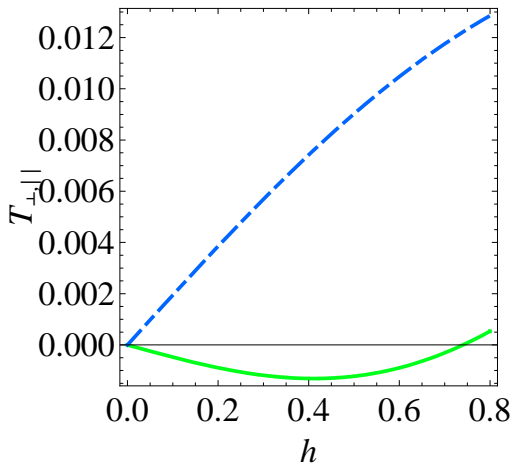}
\caption{Torkance $T_{\bot}$ (full line), $T_{||}$ (dashed line)
plotted as a function of the Zeeman energy of the fixed layer $h$. The
remaining parameters have been fixed as follows:
$\Gamma=0.5$, $r_1=r_2=r_0=0$, $\theta=\pi/2$, $k_Fd_1=k_Fd_2=3$ and $k_Fd_3=1$.}
\label{fig:fig2}
\end{figure}

In Fig.\ref{fig:fig2} the torkance components ($T_{||,\bot}$) are plotted in units of area as a function of the Zeeman energy $h$ of the fixed layer
for the remaining parameters: $\Gamma=0.5$, $r_1=r_2=r_0=0$, $\theta=\pi/2$, $k_Fd_1=k_Fd_2=3$ and $k_Fd_3=1$.
In absence of scattering potentials along the transport direction (i.e. $r_1=r_2=r_0=0$) and in presence of the exchange interaction only, the parallel component of torque satisfies the relation $T_{||}>T_{\bot}$, as in conventional spin-valves.  The component $T_{\bot}$ becomes relevant for $r_i\neq0$ (not shown here). Furthermore, if the fixed layer is made of a weak ferromagnet (i.e. $h<0.2$) the torkance components present a linear dependence with respect to $h$, while deviations from the linear behavior are observed for increasing values of the Zeeman energy. In particular, for a critical value of the Zeeman interaction, close to $h\simeq 0.75$,  the perpendicular component of the torkance $T_{\bot}$ is totally suppressed, while $T_{||}$ is the only relevant component.\\

\begin{figure}[h]
\vspace{1.0cm}
\centering
\includegraphics[scale=0.8]{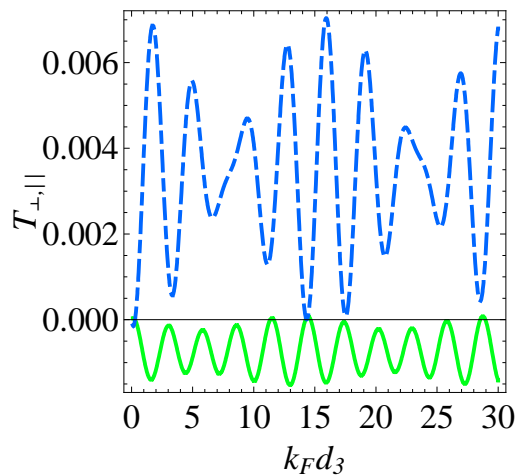}
\caption{Torkance $T_{\bot}$ (full line), $T_{||}$ (dashed line)
plotted as a function of the normalized width $k_Fd_3$ of the fixed layer F2. The
remaining parameters have been fixed as:
 $\Gamma=0.5$, $r_1=r_2=r_0=0$, $\theta=\pi/2$, $k_Fd_1=k_Fd_2=3$ and $h=0.2$.} \label{fig:fig3}
\end{figure}

\begin{figure}[h]
\vspace{1.27cm}
\centering
\includegraphics[scale=0.4]{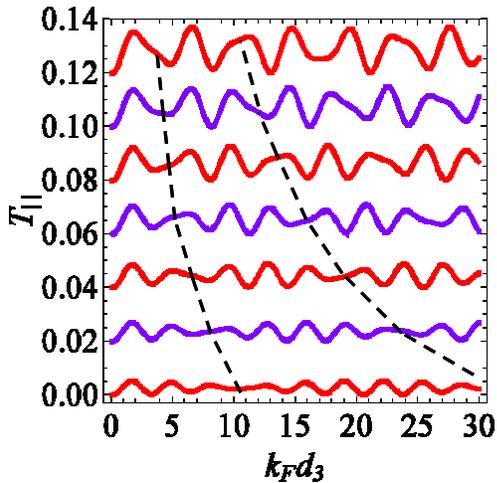}
\caption{Torkance $T_{||}$
plotted as a function of the normalized width $k_Fd_3$ of the fixed layer F2. The lower curve is plotted for $h=0.15$, while the artificial h-dependent upward shift $T_{||}(k_Fd_3) \rightarrow T_{||}(k_Fd_3)+0.02 n$, where $n=0, 1,...$  for $h=0.15, 0.2,...$ has been given to the curves to facilitate the comparison. The remaining parameters have been fixed as in Fig.\ref{fig:fig3}.}
\label{fig:fig4}
\end{figure}

\begin{figure}[h]
\vspace{1.23cm}
\centering
\includegraphics[scale=0.55]{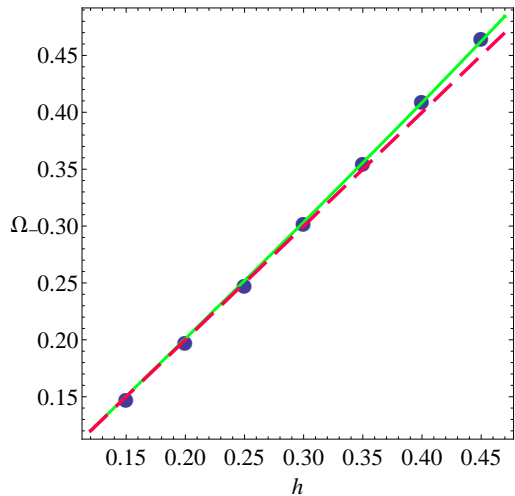}
\caption{$\Omega_{-}$ as a function of $h$ (full circles) obtained by non-linear fit procedure of the curves shown in Fig.\ref{fig:fig4}. The dashed line, inserted for comparison, represents the linear model $\Omega_-(h)=h$. The full line interpolates the computed points and shows the deviation from a simple linear model.}
\label{fig:fig5}
\end{figure}

\begin{figure}[h]
\vspace{1.0cm}
\centering
\includegraphics[scale=0.8]{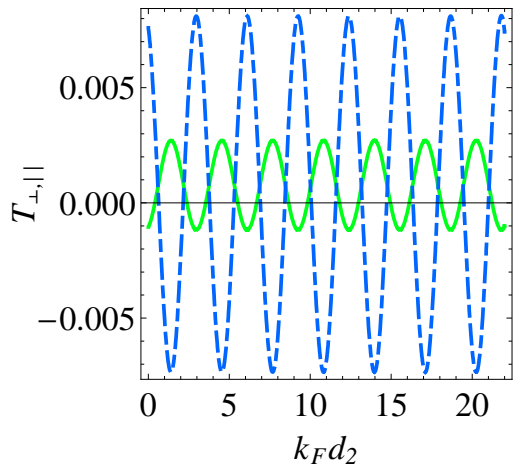}
\caption{Torkance $T_{\bot}$ (full line), $T_{||}$ (dashed line)
plotted as a function of the normalized spacer width $k_Fd_2$. The
remaining parameters have been fixed as follows:
 $\Gamma=0.5$, $r_1=r_2=r_0=0$, $\theta=\pi/2$, $k_Fd_1=4$, $k_Fd_3=2$, $h=0.25$, $\varphi=\pi/2$.}
\label{fig:f6}
\end{figure}

In Fig.\ref{fig:fig3} we plot the torkance components ($T_{||,\bot}$) as a function of the width of the fixed layer $k_Fd_3$ for
the remaining parameters: $\Gamma=0.5$, $r_1=r_2=r_0=0$, $\theta=\pi/2$, $k_Fd_1=k_Fd_2=3$ and $h=0.2$.
In agreement with what found above, in absence of scattering potentials along the $x$ direction the parallel component of the spin torque $T_{||}$
is larger than $T_{\bot}$ over a large range of the fixed layer width.
As shown in the figure, the torkance
presents a characteristic oscillatory behavior as a function of the
width of the magnetic layer.
These oscillations can be regarded as a quantum-size effect. They
reflect the perfect ballistic regime of electron transport through
the multilayer structure. The physical mechanism behind the oscillations is
the interference effect of the electrons propagating across the
non-magnetic/magnetic interface from the left lead to the right
lead and electrons propagating backwards\cite{theodonis-quantum_well_states}.  The particular value of
the oscillation follows from the values of the spin-dependent
Fermi wavevector. Furthermore, the behavior found in Fig.\ref{fig:fig3} is similar to that found in Ref.\cite{stiles02_1} (see Figs.2, 3 of the cited work)
where the torkance of a spin-valve was analyzed by \textit{ab initio} calculation.
In our case, differently from [\onlinecite{stiles02_1}],
 we observe large oscillations of the $T_{||}$ component around the mean value due to the quasi-one-dimensional character
 of our system. In the two-dimensional case, not considered here, we expect sizable changes.
 In fact, when the integration over the Fermi surface is performed taking into account all the incident directions of the electrons momentum
 a reduction of the amplitudes of the oscillating part of the torkance is expected.
Very remarkably, the oscillatory behavior shown in Fig.\ref{fig:fig3} and
particularly evident for the $T_{||}$ component displays slow and fast scales of oscillation with respect to the fixed layer width.
This can be seen by looking at the behavior of $T_{||}$ as a function of the fixed layer width $k_Fd_3$
for different values of the Zeeman energy $h$ and fixing the remaining parameters as in Fig.\ref{fig:fig3}.
The results are shown in Fig.\ref{fig:fig4}, where $T_{||}$ has been plotted vs $k_Fd_3$ for values of $h$ ranging from $h=0.15$ (lower curve)
up to $h=0.45$ (top curve) with a step of $0.05$.
The analysis of the lower curve in Fig.\ref{fig:fig4} shows an oscillating  behavior  vs $k_Fd_3$  characterized by fast oscillations with a frequency $\Omega_+$, the amplitude of the signal being modulated by a curve with frequency $\Omega_-$. As shown in Fig.\ref{fig:fig4} by increasing the Zeeman energy $h$ the larger frequency $\Omega_+$ remains almost unchanged while the smaller one $\Omega_-$ increases. This trend is represented by the dashed line in Fig.\ref{fig:fig4} .  The oscillating behavior of $T_{||}$ as a function of $k_Fd_3$ can be fitted by a non-linear regression with trial function:
\begin{eqnarray}
T_{||}(k_Fd_3)&=&T_0+\cos(\Omega_-k_Fd_3)[A\sin(\Omega_+k_Fd_3)+\\\nonumber
&+&B\cos(\Omega_+k_Fd_3)],
\end{eqnarray}
$T_0, A, B, \Omega_{\pm}$ being fitting parameters.
In Fig.\ref{fig:fig5} we report the fitting analysis of $\Omega_-(h)$ (full circles) and for comparison we also plot the line $\Omega_-(h)=h$.
The analysis of Fig.\ref{fig:fig5} shows that the slow frequency $\Omega_-$ is controlled by the Zeeman energy $h$ of the fixed layer
and thus indirectly can give a measure of the polarization of the electrons belonging to the fixed layer.
Indeed, remembering that the matching conditions on the electron wavefunctions needed to compute the S-matrix involve oscillating functions of the form $\exp(\pm i k_{\sigma}d_3)$, where  $k_\sigma=k_F\sqrt{1+\sigma h}$, one expects that the scattering matrix elements to the lowest order can be approximated via a linear combination of terms $\sin(k_{\sigma} d_3)$ and $\cos(k_{\sigma} d_3)$, or equivalently by harmonic functions of argument $(k_{\uparrow}\pm k_{\downarrow})/2$. Since the spin torque depends roughly on the scattering matrix elements squared,
 one expects that the oscillation frequencies of  $T_{||,\bot}$ vs $k_Fd_3$ are of the form $\Omega_\pm=\sqrt{1+h}\pm\sqrt{1-h}$. Assuming this relation for $\Omega_+,\Omega_-$ one reproduces exactly the result in Fig.\ref{fig:fig5} in the limit of small $h$. Moreover by considering the limit
 $h \rightarrow 0$ one expects $\Omega_-\rightarrow 0$ while only the oscillations with frequency $\Omega_+$ survive. Indeed this is found
in Fig.\ref{fig:f6} where the torque components as a function of non-magnetic layer width $k_Fd_2$  is shown. Indeed, the figure presents an oscillating pattern of period $2\pi/\Omega_+=\pi$, the modulation with smaller frequency being totally absent.

\begin{figure}[h]
\vspace{1.0cm}
\centering
\includegraphics[scale=0.8]{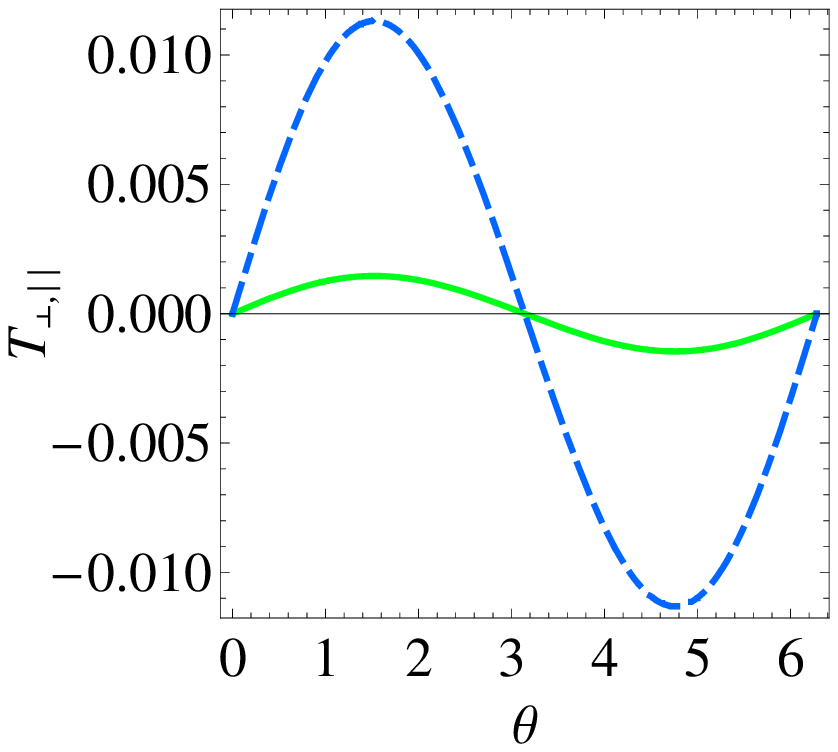}\\
\includegraphics[scale=0.8]{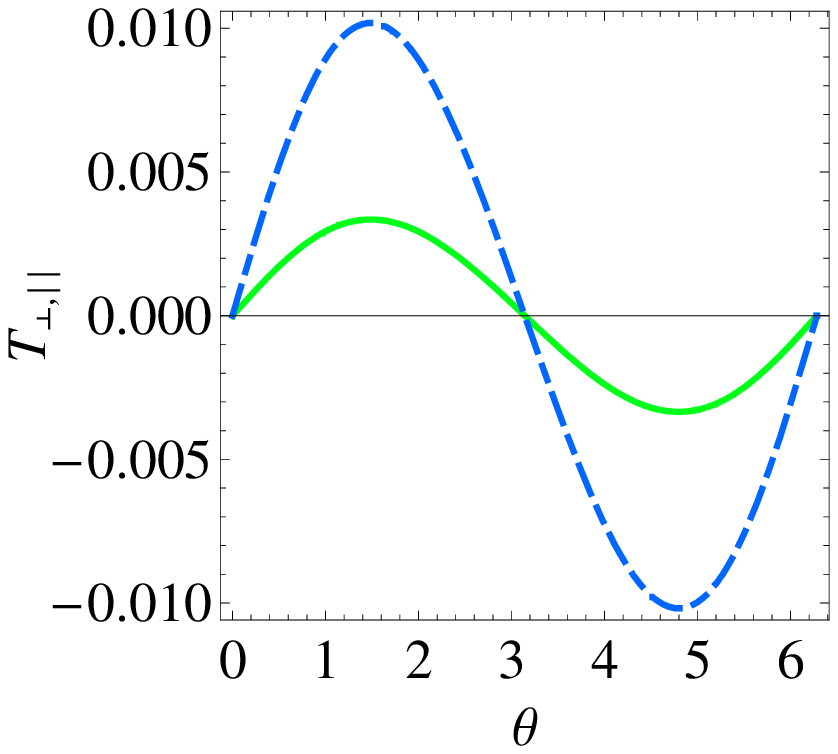}\\
\caption{Torkance $T_{\bot}$ (full line), $T_{||}$ (dashed line)
plotted as a function of the angle $\theta$. The
remaining parameters have been fixed as follows:
 $\Gamma=0.5$, $k_Fd_1=k_Fd_2=3$, $k_Fd_3=10$, $h=0.45$, $\varphi=\pi/2$ and $r_1=r_2=r_0=0$  (upper panel), or $r_1=0.1$, $r_2=0.05$, $r_0=0.2$ (lower panel).}
\label{fig:f7ab}
\end{figure}
In Fig.\ref{fig:f7ab} we show the torkance components as a function of the angle $\theta$ between the magnetizations of the regions F1 and F2. Apart from the sinusoidal behavior of the curves with respect to $\theta$, the presence of scattering potentials along the transport direction (lower panel) enhances the perpendicular component of the torkance if compared with the one obtained in the absence of scattering potentials (upper panel). Furthermore the $T_{||,\bot}$ vs $\theta$ curves present an in-phase behavior which can be altered by changing the relative size of the microstructure due to the oscillating nature of the torkance as a function of $k_Fd_3$.

\subsection{Pumping case}

Here we analyze the spin torque generated on free layer by a quantum pumping mechanism.
In particular, we adiabatically modulate in time the electrostatic potentials $r_i(t)$ as:
\begin{equation}
r_i(t)=r_i^0+r_i^{\omega}\sin(\omega t+\varphi_i),
\end{equation}
where $i=1,2$, and make the following choice for the pumping phase $\varphi_1=0,\varphi_2=\pi/2$.

\begin{figure}[h]
\vspace{1.0cm}
\centering
\includegraphics[scale=0.8]{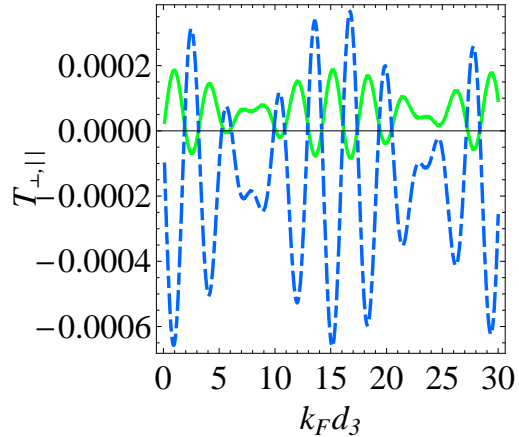}\\

\caption{Torkance $T_{\bot}$ (full line), $T_{||}$ (dashed line)
plotted as a function of the normalized fixed layer width $k_Fd_3$. The
remaining parameters have been fixed as follows:
 $\Gamma=0.5$, $r^0_1=r^0_2=r_0=0$, $r^{\omega}_1=r^{\omega}_2=0.1$, $\theta=\pi/2$, $k_Fd_1=k_Fd_2=3$, $h=0.2$, $\varphi=\pi/2$.} \label{fig:f8}
\end{figure}

\begin{figure}
\vspace{1.0cm}
\centering
\includegraphics[scale=0.8]{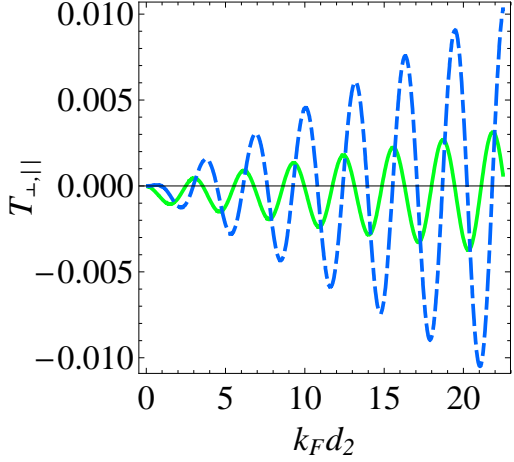}\\

\caption{Torkance $T_{\bot}$ (full line), $T_{||}$ (dashed line)
plotted as a function of the normalized spacer width $k_Fd_2$. The
remaining parameters have been fixed as follows:
 $\Gamma=0.5$, $r^0_1=r^0_2=r_0=0$, $r^{\omega}_1=r^{\omega}_2=0.1$, $\theta=\pi/2$, $k_Fd_1=4$, $k_Fd_3=2$, $h=0.25$, $\varphi=\pi/2$.} \label{fig:f9}
\end{figure}

 In Fig. \ref{fig:f8} we plot the torkance $T_{||,\bot}$ as a function of the fixed layer width $k_Fd_3$ fixing the remaining parameters as:
$\Gamma=0.5$, $r^0_1=r^0_2=r_0=0$, $r^{\omega}_1=r^{\omega}_2=0.1$, $\theta=\pi/2$, $k_Fd_1=k_Fd_2=3$, $h=0.2$, $\varphi=\pi/2$.
By comparing Fig.\ref{fig:f8} with the analogous figure obtained for the dc case (i.e. Fig.\ref{fig:fig3})
one observes that the pumping procedure modifies the sign of the mean values of torque components and simultaneously
enhances the relative amplitudes of the oscillating patterns.  As discussed above, in presence of scattering potentials along the transport direction,  we expect the perpendicular component of the spin torque to be more relevant and, compared to the dc case, this is particularly true for the pumping case where two scattering potentials are modulated in time.
Again one can distinguish two frequencies of oscillation of the torque components $\Omega_\pm$ whose values are the same obtained in the dc case. 
Apart from the changes described above,
the quantum pumping mechanism may induce relevant effects on the spin torque experienced by the free layer
due to the parametric derivatives of the scattering matrix that appear in (\ref{eq:torque_par/perp_free-ac}).
This is clearly seen in Fig.\ref{fig:f9} where the torque components $T_{||,\bot}$ are plotted as a function of the metallic spacer width $k_Fd_2$. Differently
from the dc case shown in Fig.\ref{fig:f6}, in Fig.\ref{fig:f9} we observe a linear increasing of the oscillation amplitudes
of the torque components and the lost of periodicity of $T_{||,\bot}$ vs $k_Fd_2$.
This is a {\it magnification} effect of the torque pumped in the system that can be exploited in current experiments.
Let us note that the value $k_Fd_2=20$ corresponds to a spacer width of $\sim 3.18 \lambda_F$ thus our  balistic treatment is still appropriate.
The linear increasing of the oscillation amplitude
can be naively explained observing that the pumped torque is related to the derivative of a periodic function of $k_Fd_2$,
 $F=a\sin(\sqrt{1-r_2} k_Fd_2)+b\cos(\sqrt{1-r_2} k_Fd_2)$ with respect to $r_2$ (barrier height) that gives a coefficient $k_Fd_2$.
\begin{figure}
\vspace{1.0cm}
\centering
\includegraphics[scale=0.8]{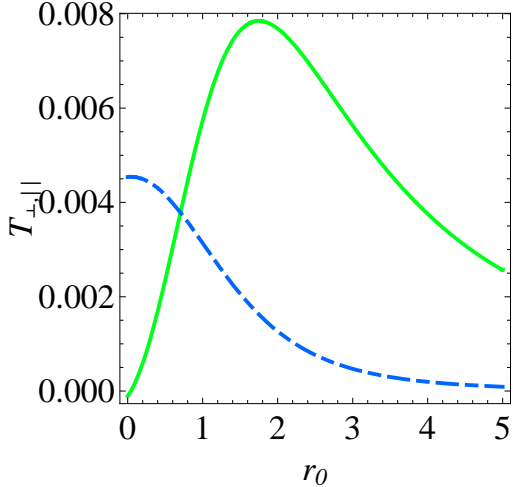}\\

\caption{Torkance $T_{\bot}$ (full line), $T_{||}$ (dashed line)
plotted as a function of $r_0$. The
remaining parameters have been fixed as follows:
 $\Gamma=0.5$, $r^0_1=r^0_2=0$, $r^{\omega}_1=r^{\omega}_2=0.1$, $\theta=\pi/2$, $k_Fd_1=4$, $k_Fd_2=10$, $k_Fd_3=2$, $h=0.25$, $\varphi=\pi/2$.} \label{fig:f10}
\end{figure}
Concerning the dependence of the torque components on the scattering potential amplitude along the transport direction, we can observe in Fig.\ref{fig:f10} an increasing of the perpendicular component of spin torque $T_{\bot}$ as a function of the barrier located on the free-layer $r_0$, while the component $T_{||}$ becomes very small. When the transparency at the free-layer $r_0$ becomes small (i.e. for high values of $r_0$) both the components of the spin torque decrease due to a suppression of the spin-fluxes.
In Fig.\ref{fig:f11} we present the torque components as a function of the Zeeman energy $h$ of the fixed layer for $k_Fd_3=6$ (upper panel) or $k_Fd_3=5$ (lower panel) and fixing the remaining parameters as follows: $\Gamma=0.5$, $r^0_1=r^0_2=0$, $r_0=2$, $r^{\omega}_1=r^{\omega}_2=0.1$, $\theta=\pi/2$, $k_Fd_1=3$, $k_Fd_2=3$, $\varphi=\pi/2$. One can notice a strong dependence on the width of the fixed layer and, in particular, one observes a sign reversal of $T_{\bot}$ at varying $h$ (lower panel). This behavior can be understood by the analytical expression of
 the torque which is an oscillating function of argument $h k_Fd_3$. As shown above a certain value of $h$, i.e. $h_c\simeq 0.75$,
 the magnetic barrier height increases reducing the transmission of spins across the layer and thus a suppression of the spin-torque is observed (over-barrier reflection).
\begin{figure}[h]
\centering
\includegraphics[scale=0.8]{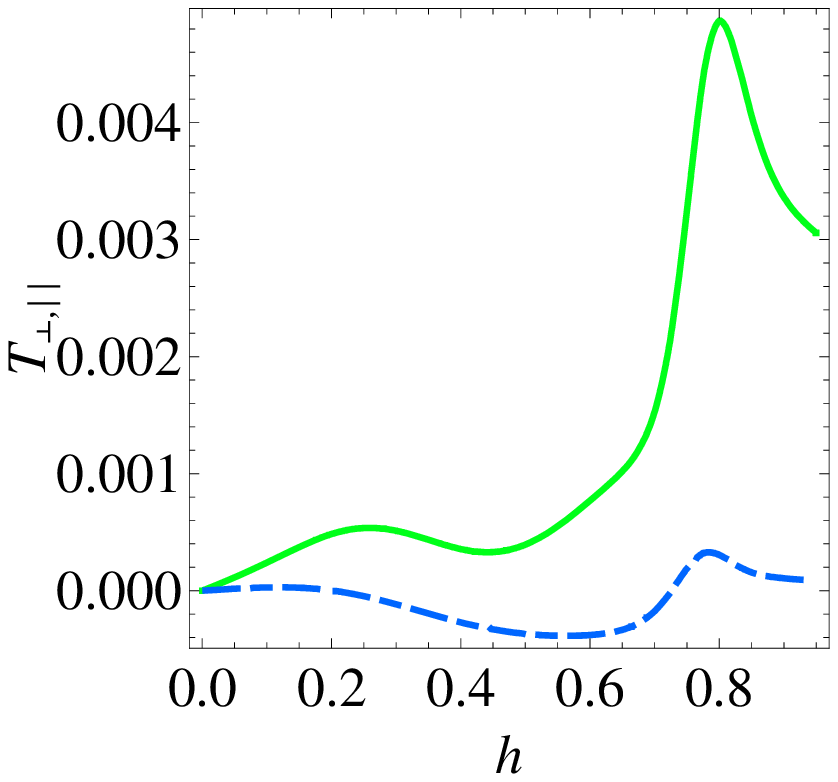}\\
\vspace{0.4cm}
\includegraphics[scale=0.8]{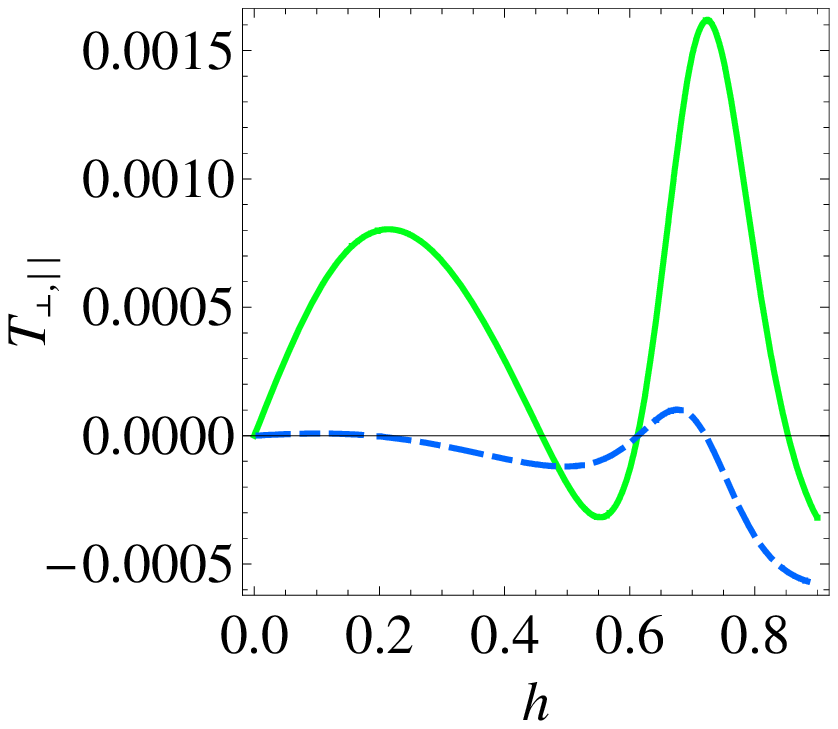}\\

\caption{Torkance $T_{\bot}$ (full line), $T_{||}$ (dashed line)
plotted as a function of the normalized Zeeman energy $h$. The
remaining parameters have been fixed as follows:
 $\Gamma=0.5$, $r^0_1=r^0_2=0$, $r_0=2$, $r^{\omega}_1=r^{\omega}_2=0.1$, $\theta=\pi/2$, $k_Fd_1=3$, $k_Fd_2=3$, $\varphi=\pi/2$, while the width of the fixed layer is set to $k_Fd_3=6$ (upper panel) or $k_Fd_3=5$ (lower panel).} \label{fig:f11}
\end{figure}

\begin{figure}[h]
\vspace{1.0cm}
\centering
\includegraphics[scale=0.8]{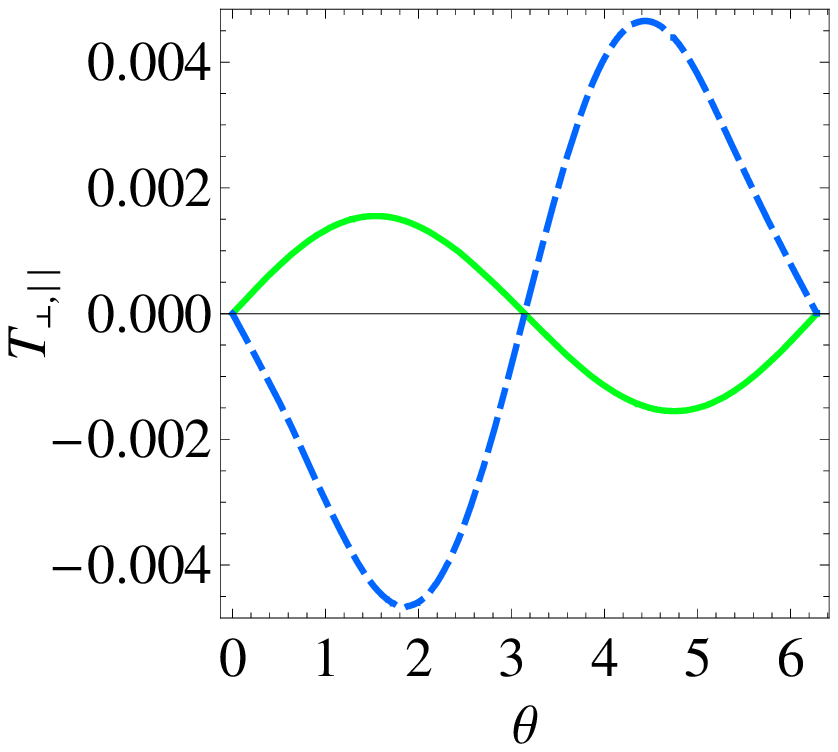}\\
\includegraphics[scale=0.75]{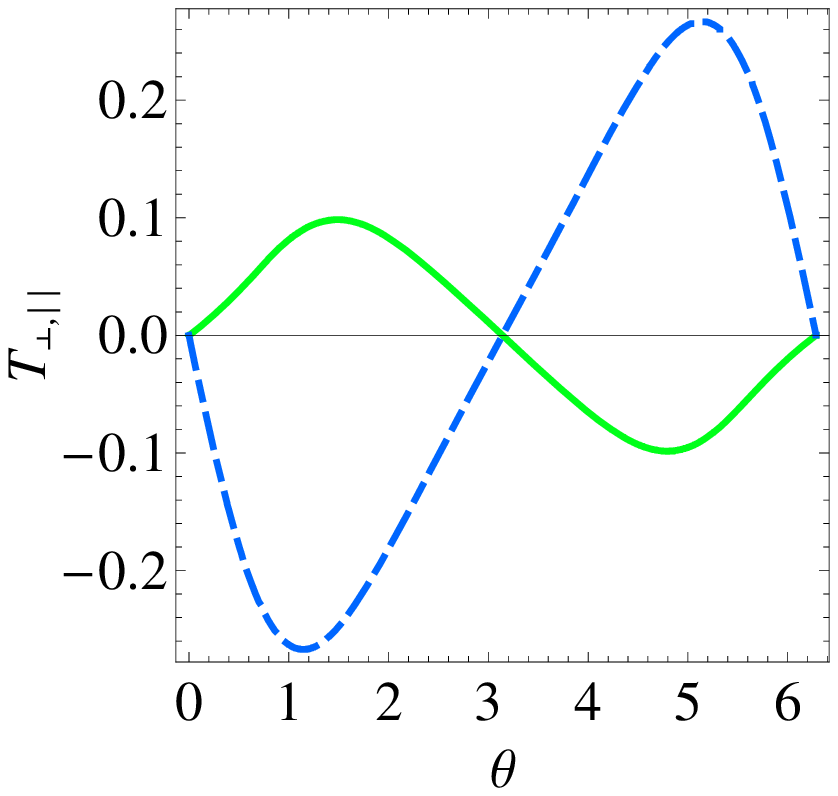}\\
\caption{Torkance $T_{\bot}$ (full line), $T_{||}$ (dashed line)
plotted as a function of the angle $\theta$. The
remaining parameters have been fixed as follows:
 $\Gamma=0.5$, $k_Fd_1=k_Fd_2=3$, $k_Fd_3=10$, $h=0.45$, $\varphi=\pi/2$, $r^{\omega}_1=r^{\omega}_2=0.3$ and $r^0_1=r^0_2=r_0=0$  (upper panel);   $\Gamma=0.65$, $k_Fd_1=3$, $k_Fd_2=15$, $k_Fd_3=10$, $h=0.85$, $\varphi=\pi/2$, $r^{\omega}_1=r^{\omega}_2=0.3$ and $r^0_1=r^0_2=r_0=0$ (lower panel).}
\label{fig:f12}
\end{figure}
In Fig.\ref{fig:f12} we report the torkance as a function of the angle $\theta$ between the two magnetizations
$\hat{n}_1$ and $\hat{n}_2$, while setting the other parameters as in the figure label.
As shown, the perpendicular component of the
torque close to $\theta=\pi/2$ is $\sim 30 \%$ of the parallel component due to the scattering potentials $r_{1,2}^\omega$, while a sinusoidal behavior with respect to $\theta$ is observed for both curves. Again one observes a magnification effect of the spin torque by increasing the width $k_Fd_2$ of the non-magnetic metallic layer between the two ferromagnetic layers F1 and F2 (the values of the torque increases by a factor $10^2$).
The magnification effect reported above can be easily understood as a combined effect originated by the pumping-induced magnification related to the increasing of the size $k_Fd_2$ of the interstitial layer further enhanced by exploiting strong polarized magnetic layers F1/F2 (i.e. by setting $h=0.85$ and $\Gamma=0.65$ instead of the values $h=0.45$, $\Gamma=0.5$ used in the upper panel of Fig.\ref{fig:f12}). It is worth to mention that the torque per unit of area induced by the quantum pumping within the weak pumping regime analyzed here is typically a fraction in unit of $\mu eV/area$. At pumping frequency of $\nu_0=300$ MHz, the energy scale normalizing the torkance $\hbar \nu_0/2$ is given by $0.1 \ \mu eV$, i.e the $ \sim 12.5\%$ of the maximum value ($\sim 0.8 \ \mu eV/area$) obtained within the dc case in Ref.[\onlinecite{kalitsov09_mtj}] by considering a MTJ. Thus in our simulations, considering $\nu_0=300$ MHz, we have verified that the maximum value of the torque obtained by a weak pumping procedure ranges from $3\%$ up to $6\%$ of the maximum value of the spin torque obtained in Ref.[\onlinecite{kalitsov09_mtj}]. Higher values of the spin torque produced by a quantum pumping mechanism can be obtained beyond the weak pumping regime.

\section{Conclusions}

We analyzed the spin-torque in a NM/F1/NM/F2/NM microstructure by a scattering matrix approach and considering different mechanisms for its generation:
1) the spin torque induced by a dc voltage $V$ applied to the whole system; 2) the spin torque activated by the quantum pumping technique. While the first method is widely studied in the present literature,  the second one based on the quantum pumping has been initially proposed in Ref.[\onlinecite{romeo_2010}] and a complete analysis has been performed in the present work. In particular, we have analyzed the quantum size effects induced by the finite width of the ferromagnetic fixed layer F2 both when the system is forced by a dc bias and in the pumping case and found interesting features related to the quantum pumping mechanism.
In the case of an external dc bias, the most evident feature is the presence of an oscillatory behavior (detectable also in the pumping case) of the $T_{||,\bot}$ vs $k_F d_3$ curves characterized by frequencies $\Omega_{\pm}$ directly related to the Zeeman energy of F2.
These oscillations reflect the
perfect ballistic regime of electron transport across the
whole system and can give important information on the polarization at the interface of the magnetic layers.
Indeed,  the spin
torque arises either as an interference effect between
spin up electrons propagating across the ferromagnetic region from the
R lead to the L lead and spin-down electrons propagating
backwards or have to be ascribed to quantum well states, i.e. to an
interference effect in a single spin channel.
The finite layer width effects described within the dc case by using our theory present qualitative agreement with recent studies on the spin torque generated in a Cu/Fe/MgO/Fe/Cu tunnel junction\cite{stiles02_1} and in a Cu/Co/Cu/Ni/Cu system\cite{carva09}.
Then, we have proposed a parametric quantum pumping of spin torque.
The pump works by means of two external gates able to produce out-of-phase voltage modulations on two nonmagnetic regions attached to the free layer (i.e. a thin ferromagnetic region F1). The underlying idea is that in a magnetic layered
structure a pumping mechanism can activate spin currents other
than charge currents and thus a spin-torque is generated on the
magnetic layer subject to the spin-current gradient.
This peculiar way of generating spin torque is strongly affected by the dependence of scattering matrix of the microstructure on the pumping parameters (i.e. the external voltages controlled by the gates G1/G2).
As a consequence of this parametric dependence of the scattering matrix, a peculiar magnification effect of the perpendicular component of torque has been predicted, the latter feature being particularly appealing to test the proposed theory using MTJs or exploiting a modified system similar to the one described in Ref.[\onlinecite{carva09}].
Indeed, by increasing the width of the nonmagnetic spacer $k_F d_2$, a magnification of the torque components has been observed, differently from the dc case where the $T_{||,\bot}$ vs $k_F d_2$ curves present a simple oscillating behavior.
 Our estimate of the spin torque induced by the weak pumping (using a pumping frequency $\nu_0=300$MHz) is $6-7\%$ of the one obtained conventionally using dc voltages, nevertheless the effects of magnification can be efficiently exploited beyond the weak pumping regime to obtain values of $T_{||,\bot}$ similar to the one observed in the dc case. Apart from the technological motivations supporting our work, the quantum pumping of spin torque can be considered as the prototype of a new class of quantum pumps able to \textit{pump a vector} (i.e. the torque) instead of a scalar (i.e. the electron/hole charge) and can be relevant to further test the quantum effects in nano-electronics.

\bibliographystyle{prsty}

\begin{thebibliography}{99}
\bibitem{note} We consider that the magnetization of the layer centered at $x=0$ is  $\mathcal{B}_1(x)$ in $[-w,w]$ and zero elsewhere. When $2w<\lambda_F$  one assumes that the electrons experience a delta-like potential whose amplitude is related to the mean magnetization $\bar{\mathcal{B}}_1=(2w)^{-1}\int_{-w}^{w}\mathcal{B}_1(x) dx$ on the free-layer. Thus, imposing that $\mathcal{B}_1(x)=B_1 \ell \delta(x)$ and by assuming that $B_1=\bar{\mathcal{B}}_1$, we obtain $\ell \approx 2w$.
\end{thebibliography}

\end{document}